\documentclass[useAMS,usenatbib,12pt,psfig]{mnras} 
\usepackage{color}
\usepackage{graphicx,amsmath}
\usepackage{rotating}
\usepackage{txfonts}
\usepackage{pdflscape,lscape}
\usepackage[T1]{fontenc}
\usepackage{aecompl}

\newcommand{\hi}{H\,{\sc i}}

\newcommand{\prin}{$^{\prime\prime}$}

\newcommand{\km}{km\,s$^{-1}$}
\newcommand{\degree}{$^{\circ}$}
\newcommand{\halpha}{H${\alpha}$}

\newcommand{\msolar}{M$_{\odot}$}
\newcommand{\mstar}{M$_{*}$}

\newcommand{\mhi}{$M_{\mathrm {HI}}$}

\newcommand{\msolaryr}{M$_{\odot}$\,yr$^{-1}$}

\newcommand{\af}{$A_\mathit{flux}$}
\newcommand{\dvs}{$\Delta_{sys}$}
\newcommand{\re}{$R_{e}$}
\newcommand{\sd}{SdI--2}

\begin{document}

\title{ \textcolor{black}{Resolved HI in two \textcolor{black}{ultra--diffuse galaxies} from  contrasting non-cluster environments} }

\author [Scott {\it{et al.}}]{T. C. Scott$^{1}$\thanks{tom.scott@astro.up.pt}, Chandreyee Sengupta,$^{2}$, P. Lagos$^{1}$, Aeree Chung$^{3}$, O. Ivy Wong$^{4,5,6}$ \\
$^{1}$ Institute of Astrophysics and Space Sciences (IA), Rua das Estrelas, 4150-762 Porto, Portugal\\
$^{2}$ Purple Mountain Observatory, No.8 Yuanhua Road, Qixia District, Nanjing 210034, China\\
$^{3}$ Department of Astronomy, Yonsei University, 50 Yonsei-ro, Seodaemun-gu, Seoul, Republic of Korea\\
$^{4}$ CSIRO Astronomy \& Space Science, PO Box 1130, Bentley, WA 6102, Australia\\
$^{5}$ ICRAR-M468, University of Western Australia, Crawley, WA 6009, Australia\\
$^{6}$ ARC Centre of Excellence for All Sky Astrophysics in 3 Dimensions (ASTRO 3D), Australia}

\date{Received  ; accepted  }
\date{}
\pagerange{\pageref{firstpage}--\pageref{lastpage}} \pubyear{}

\maketitle

\label{firstpage}

\begin{abstract}

We report on the first resolved   \hi\ observations of two blue ultra-diffuse galaxies (UDGs) using the Giant Metrewave Radio Telescope (GMRT). These observations add to the so-far limited number of UDGs with resolved \hi\ data.   The targets are from contrasting non--cluster environments: UDG--B1 is projected in the outskirts of Hickson Compact Group 25 and   Secco--dI--2 (SdI--2) is an isolated UDG. These UDGs also have contrasting effective radii with \re\ of  3.7 kpc (similar to the Milky Way) and 1.3 kpc respectively. SdI--2 has an unusually large $\frac{M_{HI}}{M_*}$ ratio =28.9, confirming a previous single dish \hi\ observation.  Both galaxies display \hi\ morphological and kinematic signatures consistent with a recent tidal interaction, which is also supported by observations from other wavelengths, including optical spectroscopy. Within the limits of the observations' resolution our analysis indicates that SdI--2  is dark matter-dominated within its \hi\ radius and this is also likely to be the case for UDG--B1.Our study highlights the importance of high spatial and spectral resolution \hi\ observations for the study of the dark matter properties of UDGs.  

\end{abstract}


\section{Introduction}
\label{intro}
\textcolor{black}{Recent studies} \citep{vdokkum15,koda15,yagi16}  have reported over 1000 extended diffuse galaxies in and surrounding the Coma galaxy cluster (z = 0.0231). These galaxies have central surface brightnesses \textcolor{black}{($\mu_g$)} of  $\sim$ 24 -- 26 mag arcsec$^{-2}$ , effective radii\footnote{The effective radius of a galaxy is the radius \textcolor{black}{at which half of the total light  is emitted}} (\re) $\sim$ 1.5 -- 5 kpc,   median stellar masses of $\sim$6 $\times$ 10$^7$ \msolar\ and were designated by \cite{vdokkum15} as \textcolor{black}{ultra--diffuse galaxies}  (UDGs). These UDGs have a \textcolor{black}{median} \re\ similar to L* spirals (\re\ $ \sim$ 3.7 kpc) but their median stellar masses are more typical of dwarf galaxies \textcolor{black}{$\sim$10$^{7-8}$ \msolar\ \citep[and references therein]{lagos11}}. While faint, extended, low surface brightness galaxies (LSBs) are not a recent discovery, the Coma UDGs reveal their higher abundance in dense environments \citep{vaderBurg17}. Compared to classical LSBs, UDGs are optically fainter and often more extended \citep{yagi16} with a wide range of optical \re. \cite{conselice18}, argues UDGs are part of an earlier reported population of Low--Mass Cluster Galaxies (LMCGs) and attributes their large diameters to interactions with the cluster environment and expected them to be dark matter (DM) dominated.  \textcolor{black}{On the face of it, the reported tight correlation between the abundance of UDGs and the mass of the host clusters and groups \citep{vaderBurg17,roman17}} supports environmentally driven formation scenarios, such as those proposed by \cite{Baushev16,yozin2015,carleton18}. \textcolor{black}{\cite{Janssens19} reported on the  abundance and distribution of UDGs  in 6 masive clusters at z = 0.308 to 0.545. The asymmmetric distribution of UDG around 4 of the \cite{Janssens19} clusters and the anticorelation between  the UDG  and compact dwarf galaxy distributions highlights the likely role of the cluster environment in both the formation and evolution of UDGs.}  However, UDGs are also found in groups and in isolation which has lead to alternative formation models which rely, to a \textcolor{black}{greater} or lesser extent, on secular processes, e.g. objects from the high--end tail of the DM halo spin parameter distribution \citep{amorisco16} or star formation (SF) driven gas outflows generating extended discs and quenching SF \citep{DiCintio17}. Whether UDG/LMCGs (referred from here on as UDGs) turn out to be a subset of classical LSBs, as suggested by \cite{Tanoglidis20}, or separate class(es)  of their own, is likely to depend on the so-far unresolved question of how LSBs and UDGs form. Attention is turning toward the DM content and its distribution in UDGs because of its critical role in the formation and evolution of their observable baryonic components. \textcolor{black}{Beyond the cluster environment, where significant ram pressure stripping of a UDGs \hi\ is not expected, \hi\ studies become feasible.}  Resolved \hi\ \textcolor{black}{in particular} is well suited to both the detection of recent interactions and determination of DM properties and as a result, can assist in answering these questions.

In this paper, we present resolved Giant Metrewave Radio Telescope (GMRT) \hi\ observations of two blue (g -- i $<$ 0.6) UDGs from contrasting non--cluster environments; UDG--B1 is projected in the outskirts of Hickson Compact Group 25 (HCG\,25) \citep{roman17,spekkens18} while  Secco--dI--2 (hereafter \sd) is a relatively isolated \textcolor{black}{blue}  UDG  \citep{bellazzini17}.  \textcolor{black}{Because of gas \textcolor{black}{stripping} in dense environments we would expect UDGs in those environments to have redder colours than the the   g -- i $<$ 0.6 colour}  predicted for UDGs in low density environments \citep{liao19} and UDG--B1 is the bluest UDG (g -- i = 0.27$\pm$ 0.05) in a sample of bluer (0.27 $<$ g -- i $<$ 0.55) UDGs surrounding Hickson Compact groups from \cite{roman17}.   Table \ref{table_1}  gives a summary of the \textcolor{black}{UDG--B1 and \sd} properties. \textcolor{black}{There are differing definitions in the literature of a UDG, in particular,  the criteria for their \re, e.g. \citep[][$>$0.7 kpc]{yagi16},   \cite[][$>$1.3 kpc]{roman17}, although the \cite{vdokkum15} definition  of $>$1.5 kpc is widely used.  \sd\ (\re\ = 1.3 $\pm$0.1)  has previously been classified as a UDG by \cite{bellazzini17,papastergis17} and \citep{wang_j20}. But we note that while the \sd\ \re\  is above the \re\ of  $\sim$ 1kpc of local group dwarf galaxies \citep{Guo20} it is below the minimum \re\ of 1.5 kpc for UDGs adopted by many authors, e.g. \cite{forbes20} who would consider \sd\   as a ``small UDG" or ``LSB dwarf" rather than a UDG. The \re\ of \sd\ lies only slightly below the \re\ 1.5 kpc, which is confirmed by the fact that it's \re\ lies within uncertainties of three of the UDGs in the \cite{Leisman17} sample of HI rich UDGs. \textcolor{black}{Additionally \re\ measurement can change significantly with the optical band \citep{2020MNRAS.494.5293F}. \sd's \re\ (measured at 1.3$\pm$0.1 kpc in i band) may well be above 1.5 kpc in g band. Thus} given its proximity to the strictest \re\ lower limit and consistency with the previous classifications of \sd\ as a  UDG, we  adopt its classification as a UDG.}  We use optical images from the Sloan Digital Sky Survey (SDSS), IAC Stripe 82 and Pan-STARRS1 \citep{Flewelling16}  as well as UV imaging from Galaxy Evolution Explorer (\textit{GALEX})  in this paper.  UDG--B1 also has an SDSS spectrum that we utilise as part of our analysis. 

Section \ref{obs} gives details of the GMRT observations,  with  observational results in section \ref{results}.  A discussion follows in section \ref{dis} with a summary and concluding remarks in section \ref{concl}. To aid comparisons in this paper we adopt  distances  to UDG--B1 and \sd\ of 88 Mpc and 40  Mpc, respectively from  \cite{spekkens18} and \cite{papastergis17}. We also adopt their angular scales of 1\,arcmin $\sim$ 25\,kpc and 1\,arcmin $\sim$ 11\,kpc, for UDG--B1 and \sd, respectively. All $\alpha$ and $\delta$ positions referred to throughout this paper are  J2000.0.

\begin{figure*}
\begin{center}
\includegraphics[ angle=0,scale=.80] {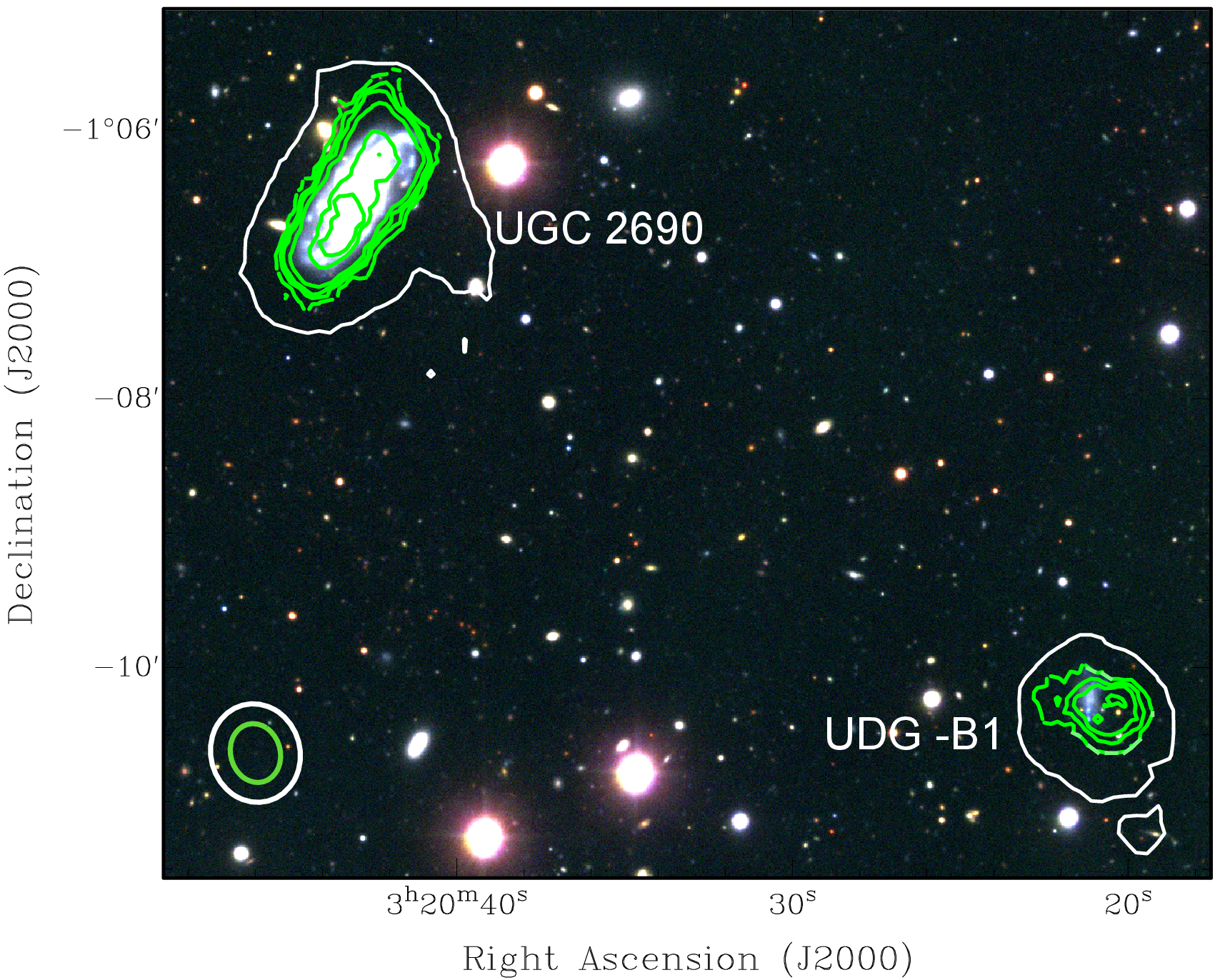}
\vspace{1cm}
\caption{
\textbf{Wide--angle region SW of HCG \,25} with  the GMRT low resolution (44.82$^{\prime\prime}$ $\times$ 40.48$^{\prime\prime}$) \hi\ contour (white) at 0.39  $\times$ 10$^{20}$ atoms cm$^{-2}$ and medium resolution (26.46$^{\prime\prime}$ $\times$ 22.65$^{\prime\prime}$)  \hi\ contours (green) at 1.46, 2.73,  3.6, 5.5 11.0 and 13.8 $\times$ 10$^{20}$ atoms cm$^{-2}$. These contours are overlaid on an IAC stripe 82 composite g,r, i image showing \hi\ detections in UGC\,2690 and UDG\,--B1. The ellipses at the bottom left  indicate the size and orientation of the GMRT synthesized beams.  }
\label{fig1}
\end{center}
\end{figure*}

\begin{table}
\centering
\begin{minipage}{190mm}
\caption{\textcolor{black}{Properties of UDG--B1 and \sd }}
\label{table_1}
\begin{tabular}{llrr}
\hline
\textbf{Property}\footnote{From the literature as noted or if in bold typeface from this work.}&\textbf{Units}&\textbf{UDG--B1}&\textbf{ \sd}  \\ 
V$_{optical}$\footnote{\sd\  from \cite{bellazzini17}.}&[\km]& 6445$\pm$12&2549$\pm$50 \\
\textbf{V$_{HI}$ $_{W20}$}&[\km]&\textbf{6450$\pm$\,\,7}&\textbf{2565$\pm$\,\,4}\\
\textbf{W$_{W20}$}&[\km]&\textbf{90$\pm$\,15}&\textbf{80$\pm$\,\,6}\\
RA\footnote{\textcolor{black}{RA and DEC positions for UDG--B1are from \cite{roman17}\\ 
and \sd\ are from \citep{bellazzini17}.  }} &[h:m:s]&03 20 21.1&11 44 33.8\\
DEC&[d:m:s]&-01 10 \textcolor{black}{12.}0&-00 52 00.9\\
Distance\footnote{\textcolor{black}{See section 1}.}&[Mpc]&  88 & 40$\pm$1\\
\re\footnote{UDG--B1 from \cite{roman17}, \sd\ from\\ \cite{bellazzini17}. }& [kpc]&3.7$\pm$0.4&1.3$\pm$\textcolor{black}{0.1}\\
$\mu_g$(0)\footnote{UDG--B1 from \cite{trujillo2017a}.}& [mag arcsec$^2$]&24.0&$>$24.0\\
Morphology&&UDG&UDG \\
12+log(O/H)\footnote{For UDG--B1 from a SDSS spectrum of the brightest \\SF knot see Table \ref{table3} and \sd\  from \citep{bellazzini17}.} & &7.91$\pm$ 0.32 &8.1$\pm$0.2  \\

 \textbf{S$_{HI}$ (GMRT)}&[Jy \km]&\textbf{0.8$\pm$0.04}&\textbf{0.69$\pm$0.03}\\
$M_{HI}$\textcolor{black}{(GMRT)} &[10$^{8}$ \msolar]&\textbf{15}&\textbf{2.6}\\
$M_*$\footnote{UDG--B1 from \cite{roman17}, \sd\ from\\ \cite{papastergis17}.}   &[10$^{8}$ \msolar] &  2.2$\pm$0.3&0.09  \\
g -- i\footnote{UDG--B1 from \cite{roman17}. \textcolor{black}{\sd\  from \\SDSS photometry.}} &&0.27$\pm$ 0.05&\textcolor{black}{0.19}\\
\hline
\end{tabular}
\end{minipage}
\end{table}

\section{Observations}
\label{obs}
UDG--B1 and \sd\ were observed in \hi\  using the GMRT on 2018 March 8th and 9th and selected observational parameters are detailed in Table \ref{table2}. \textcolor{black}{\hi\ in UDG--B1 has previously been detected within a VLA D--configration  field  \citep{spekkens18, Borthakur10}  centred on the HCG\,25 group. The UDG--B1 \hi\ was unresolved in the VLA 70 $\times$ 50 arcsec FWHM synthised beam, which provided the motivation to \textcolor{black}{reobserve} UDG--B1 with the GMRT to obtain the first ever resolved mapping of its \hi. } The GMRT data was reduced and analyzed using the standard reduction procedures with the Astronomical Image Processing System (\textsc{aips}) software package. The flux densities are on the scale of \cite{baars77}, with uncertainties of  $\sim$ 5\%. After calibration and continuum subtraction in the uv domain the \textsc{aips,} task \textsc{imagr} was used to convert the uv domain data to \hi\ image cubes. Finally integrated \hi\ \textcolor{black}{and}  velocity field  maps were extracted from the image cubes using the \textsc{aips} task \textsc{momnt}. To study the  \hi\ distribution in detail,  image cubes with different resolutions were produced by applying different `tapers' to the data with varying uv limits. Details of the final low, medium and high  resolution maps  are given in Table \ref{table2}.

\begin{table}
\centering
\begin{minipage}{110mm}
\caption{GMRT observational and map parameters}
\label{table2}
\begin{tabular}{ll}
\hline
\textbf{UDG-B1}\\
Rest frequency & 1420.4057 MHz \\
Observation Date &2018 Mar 9\\
Integration time  & \textcolor{black}{10.0 hrs } \\
primary beam & 24\arcmin ~at 1420.4057 MHz \\
Low resolution beam--FWHP  &44.82$^{\prime\prime}$ $\times$ 40.48$^{\prime\prime}$, PA = 14.03$^{\circ}$ \\
Medium resolution beam--FWHP& 26.46$^{\prime\prime}$ $\times$ 22.65$^{\prime\prime}$, PA =  15.80$^{\circ}$ \\
High resolution  beam--FWHP & 22.73$^{\prime\prime}$ $\times$ 15.31$^{\prime\prime}$, PA = -9.61$^{\circ}$ \\
RA (pointing centre)&\textcolor{black}{ 03$^{\rm h}$ 20$^{\rm m}$ 21.102$^{\rm s}$  }\\
DEC (pointing centre)& \textcolor{black}{-- 01$^\circ$ 10$^\prime$ 13.992$^{\prime\prime}$}\\
\hline
\textbf{\sd }\\
Rest frequency & 1420.4057 MHz \\
Observation Date &2018 Mar 8\\
Integration time  & \textcolor{black}{10.0 hrs } \\
primary beam & 24\arcmin ~at 1420.4057 MHz \\
Low resolution beam--FWHP  &39.95$^{\prime\prime}$ $\times$ 35.38$^{\prime\prime}$, PA = -- 10.63$^{\circ}$ \\
Medium resolution beam--FWHP& 24.50$^{\prime\prime}$ $\times$ 21.99$^{\prime\prime}$, PA =  21.18$^{\circ}$ \\
RA (pointing centre)&\textcolor{black}{ 11$^{\rm h}$ 44$^{\rm m}$ 33.802$^{\rm s}$  }\\
DEC (pointing centre)& \textcolor{black}{-- 00$^\circ$ 52$^\prime$ 00.913$^{\prime\prime}$}\\
\hline
\end{tabular}
\end{minipage}
\end{table}

\section{Observational Results} 
\label{results}

\subsection{HI morphology  and mass} 
\label{res-morph}

UDG--B1 is projected  $\sim$ 9 arcmin (225 kpc) SW of the Hickson Compact Group, HCG\,25  centre, see Figure \ref{fig1}. \textcolor{black}{The UDG's \textcolor{black}{heliocentric radial velocity (V$_{opt}$) is} 89 \km\ higher than the mean  velocity}  \textcolor{black}{ for  HCG\,25,  V$_{opt}$ = 6356 \km\ (NED) with a dispersion ($\sigma_v$) = 61.3 \km\ \citep{Tovmassian99}}.  UDG-B1 is projected within the typical group R$_{200}$ of 500 kpc \citep{roman17}\textcolor{black}{, where R$_{200}$ is the  radius enclosing an overdensity of $>$ 200 with respect to the critical density of the universe.} Its projected position and velocity  are therefore consistent with UDG--B1 being an outskirts member of HCG\,25 \textcolor{black}{so  past interactions between the UDG--B1 and HCG\,25 group members cannot be ruled out}. \hi\ was detected in UDG-B1 with the GMRT at a velocity\footnote{The \hi\ velocity was determined from the \textcolor{black}{mid point} in the W$_{20}$ velocity range, i.e. V$_{\mathrm {HI}}$ $_{W 20}$ as definded in \cite{Reynolds20}.} (V$_{\mathrm {HI}}$) = 6450$\pm$7  \km\  and in \ HCG\,25 member UGC\,2690 (V$_{opt}$ = 6285 \km, \textcolor{black}{W$_{20}$ = 340$\pm$ 8 \km}), projected  6.6 arcmin  NE and within 165 \km\ of UDG--B1, see Figure \ref{fig1}.  Further details of the \hi\ detected in UGC\,2690 are presented in Appendix A. 

The left--hand panels of Figure \ref{fig2} show  (top to bottom) the low (44.82$^{\prime\prime}$ $\times$ 40.48$^{\prime\prime}$), medium (26.46$^{\prime\prime}$ $\times$ 22.65$^{\prime\prime}$) and high (22.73$^{\prime\prime}$ $\times$ 15.31$^{\prime\prime}$)  resolution  GMRT UDG--B1  integrated \hi\ map contours   overlaid on  an IAC Stripe 82 g, r, i composite image.  
 \textcolor{black}{The high resolution \hi\ map contours  reveal significant asymmetry in} the \hi\ morphology,  with the bulk of the high column density \hi\ offset by  $\sim $ 20 arcsec (8 kpc) from  optical centre. For UGC--B1 the GMRT flux density   (S$_{HI}$) = 0.8 Jy \km\ which \textcolor{black}{converts} to an  \mhi\ $\sim$ 1.5 $\times$ 10$^9$ \msolar. \cite{spekkens18} \textcolor{black}{have previously} reported   Green Bank Telescope (GBT) single dish and Very Large Array (VLA)  \hi\ detections for UGC--B1 with S$_{HI}$ = 0.6 $\pm$0.1 Jy \km\ from the VLA data.  At the adopted distance of 88 Mpc this VLA S$_{HI}$ implies \mhi\ = 1.1 $\times$ 10$^9$ \msolar, in good agreement with the \mhi\ derived from the GMRT. The  VLA flux density sensitivity was $\sim$ 0.5 mJy beam$^{-1}$ \citep{Borthakur10}, quite similar to GMRT medium and high resolution maps. For the GBT spectrum \cite{spekkens18} reported  a  higher  peak signal to  noise ratio (15.2)  than from the VLA spectrum (9.2). At first sight this may seem like some flux loss for the interferometric observations. However, while the GBT pointing was at the position UDG--B1  the GBT FWHP beam is $\sim$ 9 arcmin and thus the spectrum was contaminated by emission from UGC 2690.  UGC\,2690 is projected 6.6 arcmin from UDG--B1, so lies just beyond  the radius of GBT FWHP beam.  \textcolor{black}{Combining} the GMRT \hi\ spectra of UDG--B1 and  UGC\,2690, attenuated to 40\% of its actual flux, produces a spectrum consistent with the GBT spectrum.  So, for the analysis in this paper we adopt the GMRT \hi\ mass. 
Based on this \hi\ mass and the $M_*$ from Table \ref{table2} the $\frac{M_{HI}}{M_*}$ ratio for UGC--B1 is 6.8.

For \sd, Figure \ref{fig3} (Top row) shows the low (39.95$^{\prime\prime}$ $\times$ 35.38$^{\prime\prime}$) and  medium (24.50$^{\prime\prime}$ $\times$ 21.99$^{\prime\prime}$)  resolution  GMRT  integrated \hi\  map contours   overlaid on a smoothed Pan--STARRS  g, r, i composite image. Figure \ref{fig3} shows the main body of the galaxy has an \hi\  extent of $\sim$ 1.2 arcmin (13.6 kpc). A detached \textcolor{black}{\hi\ tail like  extension } was also detected to the NE of the main \hi\ body reaching $\sim$ 2.4 arcmin (26 kpc) NE of the optical centre in the  medium resolution \hi\ map. \hi\ in this extended region has a  column density maximum  $>$ 7 $\times$ 10$^{19}$ atoms cm$^2$.  
The main \hi\ body of \sd\ has optical, NUV and FUV (GALEX) and NIR (WISE 3.4 $\mu$m and 4.6 $\mu$m) counterparts, see Figure \ref{fig3} -- Lower right panel. For \sd\ the GMRT  S$_{HI}$ = 0.69$\pm$0.03 J \km\ which at the adopted distance of  40 Mpc gives  \textcolor{black}{\mhi\ $\sim$ 2.6 $\times$ 10$^8$ \msolar.} A previous \hi\ detection for \sd\ using the Effelsberg 100 m single dish telescope gave S$_{HI}$ = 0.63 Jy \km\   which converts to  \mhi\ = 2.4 $\times$ 10$^8$ \msolar\ \citep{papastergis17}, in good agreement with the \mhi\ derived from the GMRT.  For the analysis in this paper we adopt the \textcolor{black}{GMRT}  \hi\ mass. 
Based on the \textcolor{black}{GMRT} \hi\ mass and the $M_*$ from \citep{papastergis17}, the $\frac{M_{HI}}{M_*}$ ratio for \sd\  is  \textcolor{black}{28.9}, which is much higher than is typical for UDGs irrespective of \re, see Figure \ref{fig5}. As reported in \cite{papastergis17} this high  $\frac{M_{HI}}{M_*}$ ratio is consistent with the trend for  isolated UDGs to show significantly higher $\frac{HI}{M_*}$ ratios than UDGs in denser environments. \textcolor{black}{It should be remembered that this plot only considers relatively isolated \hi\ UDGs and does not include cluster galaxies which are expected to be \hi\ deficient. That said we would expect the higher \re\  region of the plot to favour high DM halo spin \citep{amorisco16} or star formation SF driven extended disc  \citep[e.g.][]{DiCintio17} formation models. Conversely,  environment density  driven formation scenarios should favour the lower \mhi/\mstar\ ratio regions of the plot, but more \hi\ observations of cluster galaxies are needed to explore this parameter space.   } 
A  NED  \textcolor{black}{search revealed no companions projected within a radius of 30 arcmin ($\sim$ 330 kpc) and a velocity range of $\pm$ 500 \km. However, the \sd\ head and tail  \hi\ morphology is suggestive of a recent interaction. Given this and that the nearerst galaxy cluster (ZwCl 1141.7-0158B)  is projected $\sim$ 900 kpc and -1055 \km\ away, we can reasonably rule out ram pressure stripping as the origin of the one--sided \hi\ tail. This  leaves an interaction with a low mass satellite or gas cloud within 7 $\times$ 10$^8$ yr} as a likely explanation for the \hi\ detached region. The above timescale is based on  \cite{holwerda11}, who predict the \hi\ perturbations from even major mergers will only remain detectable in \hi\ for a maximum of 0.4 to 0.7 Gyr. In summary the \textcolor{black}{\hi} morphologies of both UGG--B1 and \sd\ indicate a recent interaction.  The evidence for and against a recent interactions  is discussed further in Section \ref{dis_interaction}.

\subsection{HI kinematics}
\label{results_kin}
\begin{figure*}
\begin{center}
\includegraphics[ angle=0,scale=.75] {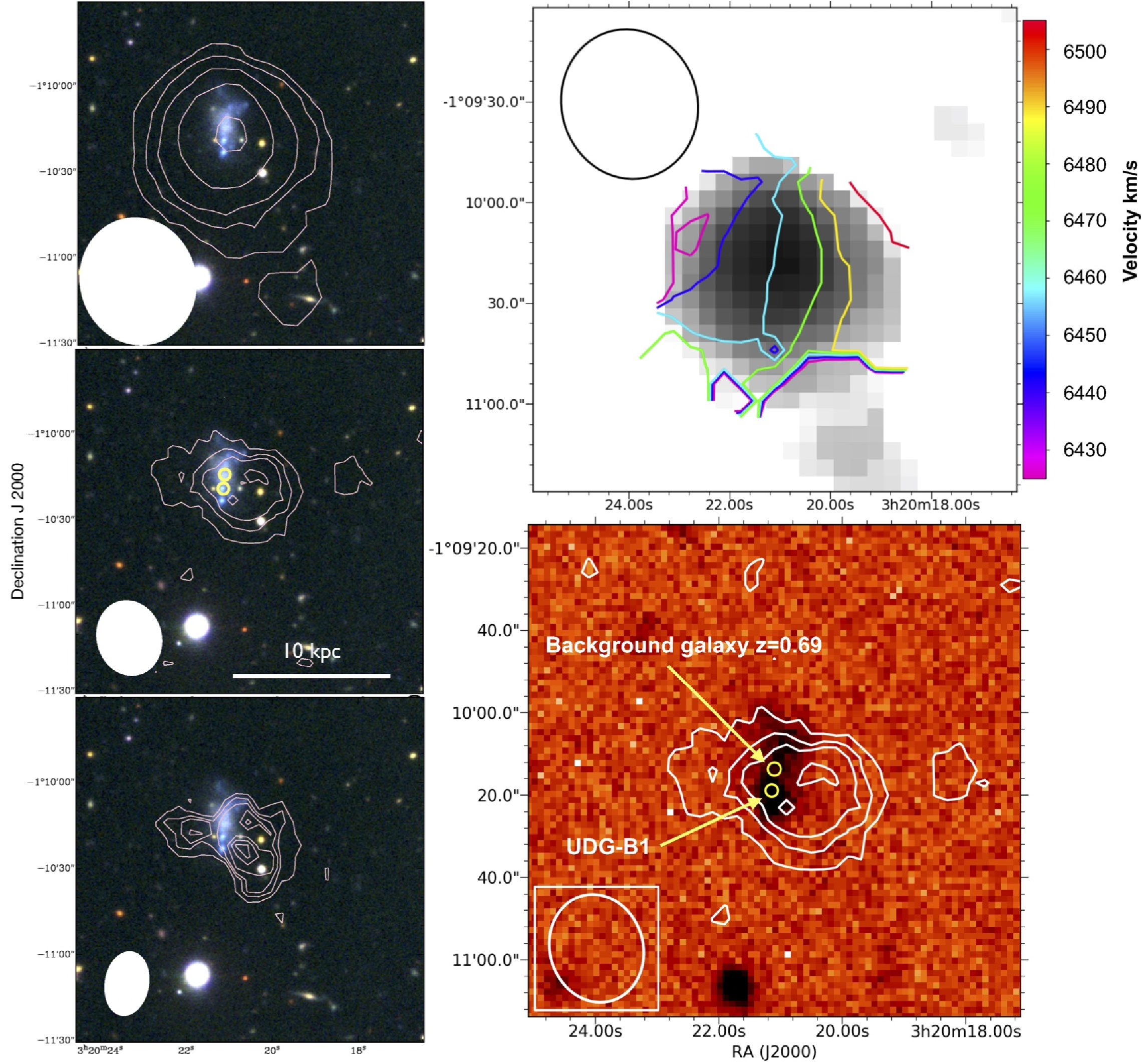}
\vspace{1cm} 
\caption{\textbf{
UDG-B1: GMRT \hi} \textit{Left column (top to bottom):}  Integrated \hi\  contours from the low (44.82$^{\prime\prime}$ $\times$ 40.48$^{\prime\prime}$), medium 26.46$^{\prime\prime}$ $\times$ 22.65$^{\prime\prime}$) and high (22.73$^{\prime\prime}$ $\times$ 15.31$^{\prime\prime}$) resolution  maps overlaid on  an IAC stripe 82 g, r, i composite   images. The column density map contours are at, low resolution;   0.39, 0.78, 1.17, 2.10 and  3.00 $\times$ 10$^{20}$ atoms cm$^{-2}$, medium resolution;  1.46, 2.73,  3.6 and 5.5 $\times$ 10$^{20}$ atoms cm$^{-2}$, high resolution; 1.26, 2.21, 2.84, 3.79, 4.74 and 5.60 $\times$ 10$^{20}$ atoms cm$^{-2}$. \textit{Right top:} \hi\ velocity field  from the low  resolution cube, with the iso--velocity \textcolor{black}{contours values, separated by 10 \km, indicated by} the colour scale.  \textit{ Lower right:}  GMRT  medium resolution (26.46\prin\ $\times$ 22.65\prin)  integrated \hi\ map contours on a smoothed GALEX NUV  image. The small  yellow circles are the positions and fields of view  of the two available SDSS spectra fibres.  The  ellipses in each panel indicates the size and orientation of the GMRT synthesized beams. }
\label{fig2}
\end{center}
\end{figure*}

Based on UDG--B1's   GMRT  \hi\ spectrum, its  V$_{\mathrm {HI}}$  = 6450 $\pm$7 \km\ and W$_{20}$ = 90 $\pm$ 15 \km. This velocity agrees within the uncertainties with the  V$_{HI}$ = 6440$\pm$5 \textcolor{black}{\km} and W$_{50}$ = 50$\pm$10 \km\ from the VLA \citep{spekkens18}.  \textcolor{black}{Figure \ref{fig2} (top right  panel) shows the UDG--B1 \hi\ low  resolution velocity field, with \hi\  detected in the velocity range 6404 \km\ to 6512  \km. } The \hi\ velocity field shows a NE to \textcolor{black}{W} gradient, which indicates that the  \hi\  disc major axis is  approximately perpendicular to the \textcolor{black}{N -- S} optical axis and \textcolor{black}{the change in the position angle of the iso--velocity  contours from the NE to W suggests a } warping of the disc, although with the caveat that the spatial resolution is low. 
We attempted to determine the rotational velocity (V$_{rot}$) from a  3D  model fit to the UDG-B1 \hi\ emission in the medium resolution cube using  \textsc{bbarolo}  \citep{DiTeodoro15}. Unfortunately, the small number of beams across the \hi\ disc  and the low \textcolor{black}{ signal to noise ratio (S/N)} meant  this attempt failed. Instead  using  inclination (i =\textcolor{black}{ 17.6}\degree) and W$_{20}$ derived from the medium resolution \hi\ cube and  equation \ref{eqn1} we estimated  V$_{rot}$ = \textcolor{black}{148} \km.

\begin{equation}
\label{eqn1}
v_{rot} =\frac{1}{2} \frac{W_{20}}{sin(i)}  
\end{equation}
where,
\[
\mathit 
sin(i) = \sqrt{ \frac{1- \left(\frac{b}{a} \right)^{\!\!2}}{1- q_0^{\!\!2}} }
\]
 a, b are the lengths of the major and minor \hi\ axes, 43.3 arcsec and 41.46 arcsec respectively from the medium resolution \hi\ map and q$_0$ =0.3 is the intrinsic axial ratio for an edge on galaxy \citep{sanchez_j10} and
 \[
\mathit 
W_{20} = \sqrt{ w_{20 raw}^2 -\delta_s^2 }
\]
where $w_{20  raw}$ is the W$_{20}$ measured from the spectrum and $\delta_s$ is a correction for instrumental broadening and turbulence (10 \km). 

However,  this inclination corrected V$_{rot}$ estimate of 148 \km\ is anomalously high by a factor of 2 to 3 for a  UDG of its \hi\ mass, even allowing for the significant uncertainties in the \hi\ major and minor axes because synthised beam is large relative to  axes. So, as a check we also calculated  V$_{rot}$ using Equation \ref{eqn1} but replacing the \hi\  axial ratio (b/a)  with the optical axial ratio of 0.46  from \cite{roman17}. This substitution  resulted in a revised  inclination of 68.6\degree\ and V$_{rot}$ = 48 \km. A visual assessment of the medium resolution \hi\ map in Figure \ref{fig1} indicates the \hi\ axial ratio is significantly lower than its optical counterpart. \textcolor{black}{It is also important to note the} calculations above assume that the \hi\ is dynamical equilibrium,  but if UDG--BI has suffered  a recent  interaction, as the analysis \textcolor{black}{of the \hi\  morphology } suggests, the apparently large \hi\  V$_{rot}$ of 148 \km\  is more likely an interaction induced  kinematic  artefact rather than an usually large \hi\ rotation velocity. The implications of the two V$_{rot}$ estimates for calculation of UDG--B1's dynamical mass are discussed in section \ref{dis_dyn}.   
 
\begin{figure*}
\begin{center}
\includegraphics[ angle=0,scale=.75] {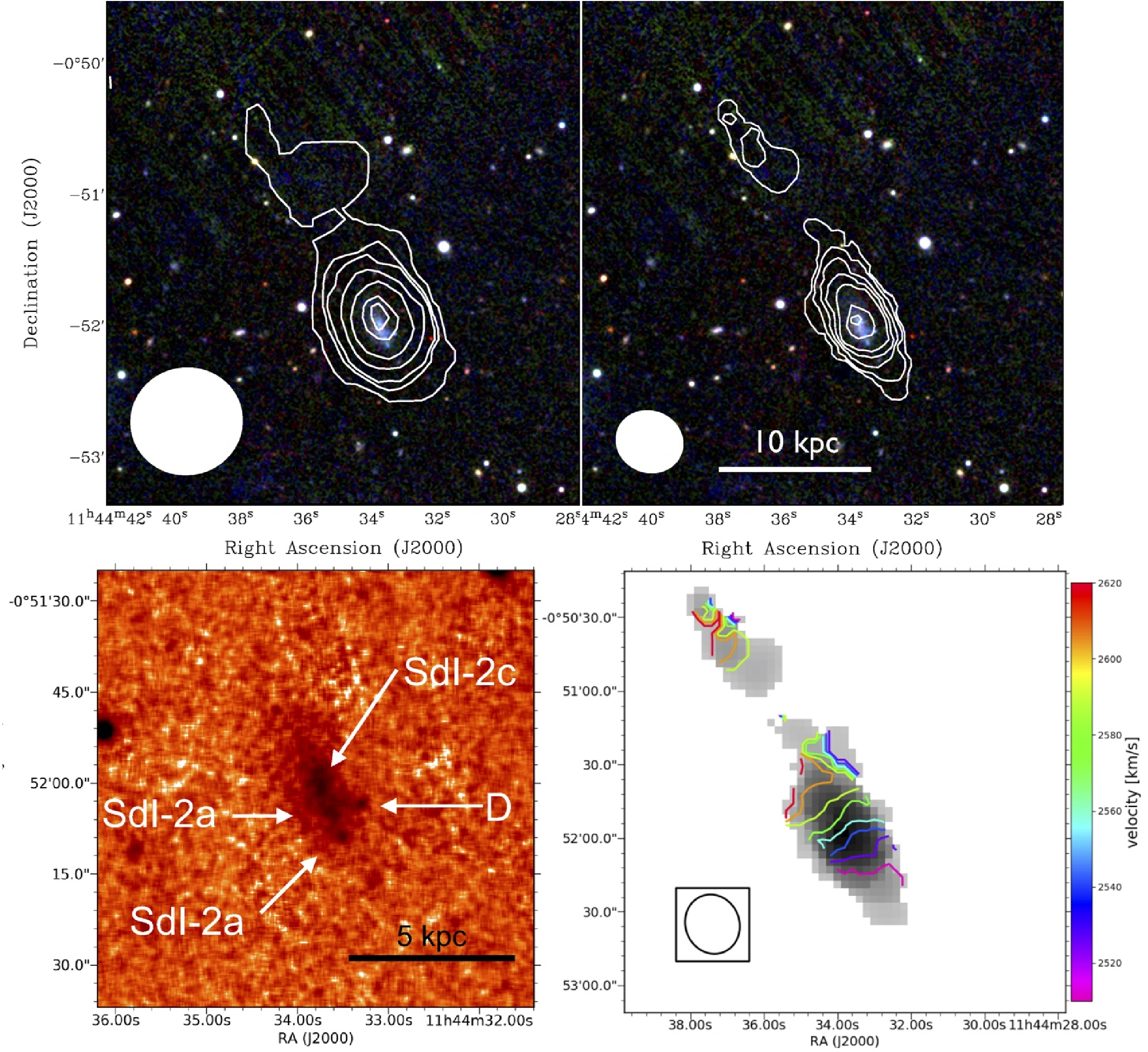}
\vspace{1cm}
\caption{\textbf{\sd:} \textit{ Top (left to right)t:}  GMRT low(39.95\prin\ $\times$ 35.38\prin) and medium (24.50\prin\ $\times$ 21.99\prin) resolution \hi\ contours overlaid on a Pan--STARRS  g,r,i--band composite  image. The column density map contours are at, low resolution;   0.15, 0.54, 0.77, 1.55, 2.33, 3.11 and 3.50 $\times$ 10$^{20}$ atoms cm$^2$, medium resolution;  0.40, 1.02, 1.83, 3.06, 4.08, 6.10 and 7.14 $\times$ 10$^{20}$ atoms cm$^2$.
 \textit{Lower left:} \sd\  smoothed NUV (GALEX)  image  with the star forming regions from Bellazzini et al. (2017) indicated. \textit{Lower right:} Medium resolution \hi\ velocity field  iso--velocity contours separated by  10 \km\ overlaid on \hi\ integrated medium resolution map.The GMRT beam sizes and orientations  are  shown with  ellipses at the bottom of each panel.  }
\label{fig3}
\end{center}
\end{figure*}

\begin{figure*}
\begin{center}
\includegraphics[ angle=0,scale=.45] {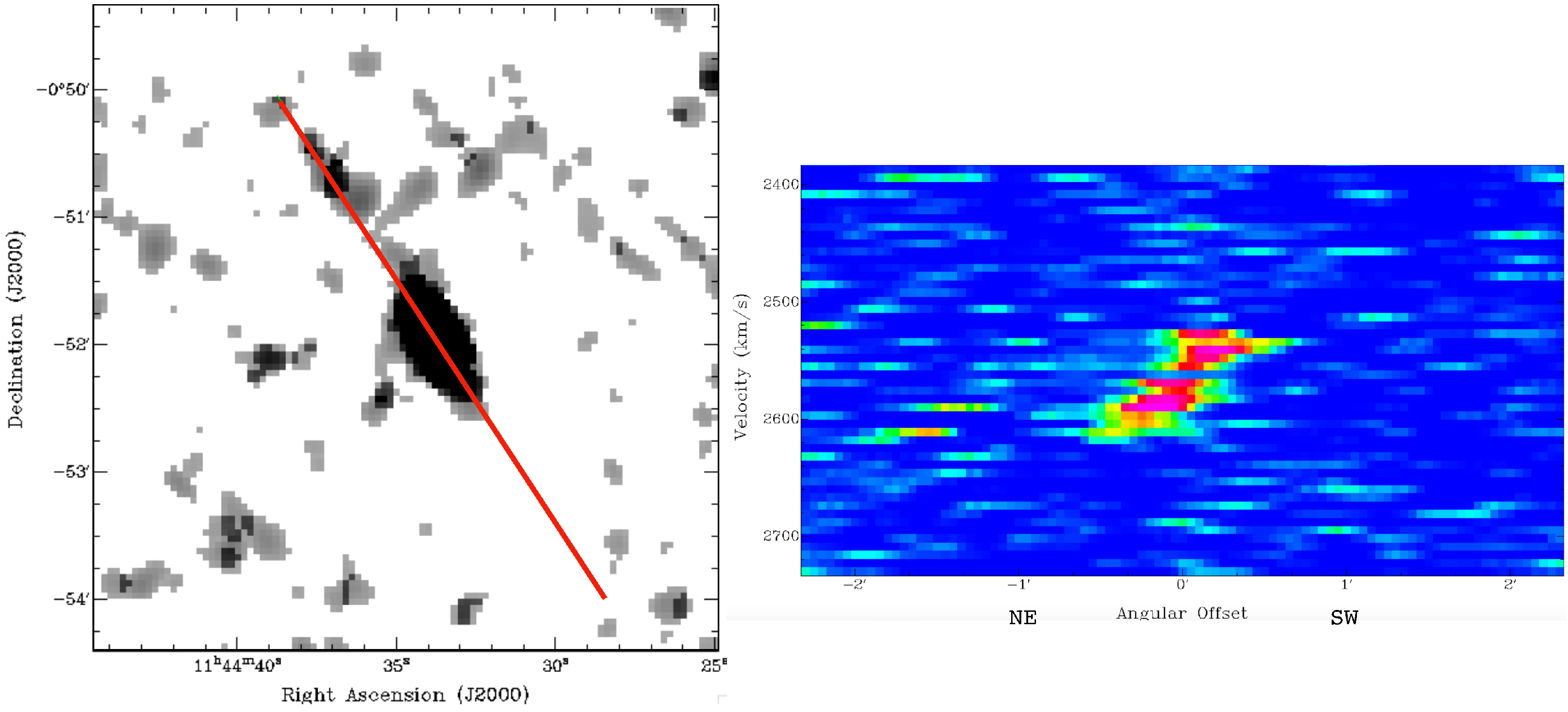}
\vspace{1cm}
\caption{\textbf{\sd\ PV diagram:} from the low resolution GMRT cube (spatial resolution $\sim$ 40 arcsec). The cut (PA 23.3\degree) is through the centre along the major \hi\ axis. Positive angular offsets are to the SW of the optical centre and negative offsets are to the NE and include the \hi\ NE detached feature. The velocity resolution of the cube is   7 \km.}
\label{fig6}
\end{center}
\end{figure*}

\begin{figure*}
\begin{center}
\includegraphics[ angle=0,scale=.95] {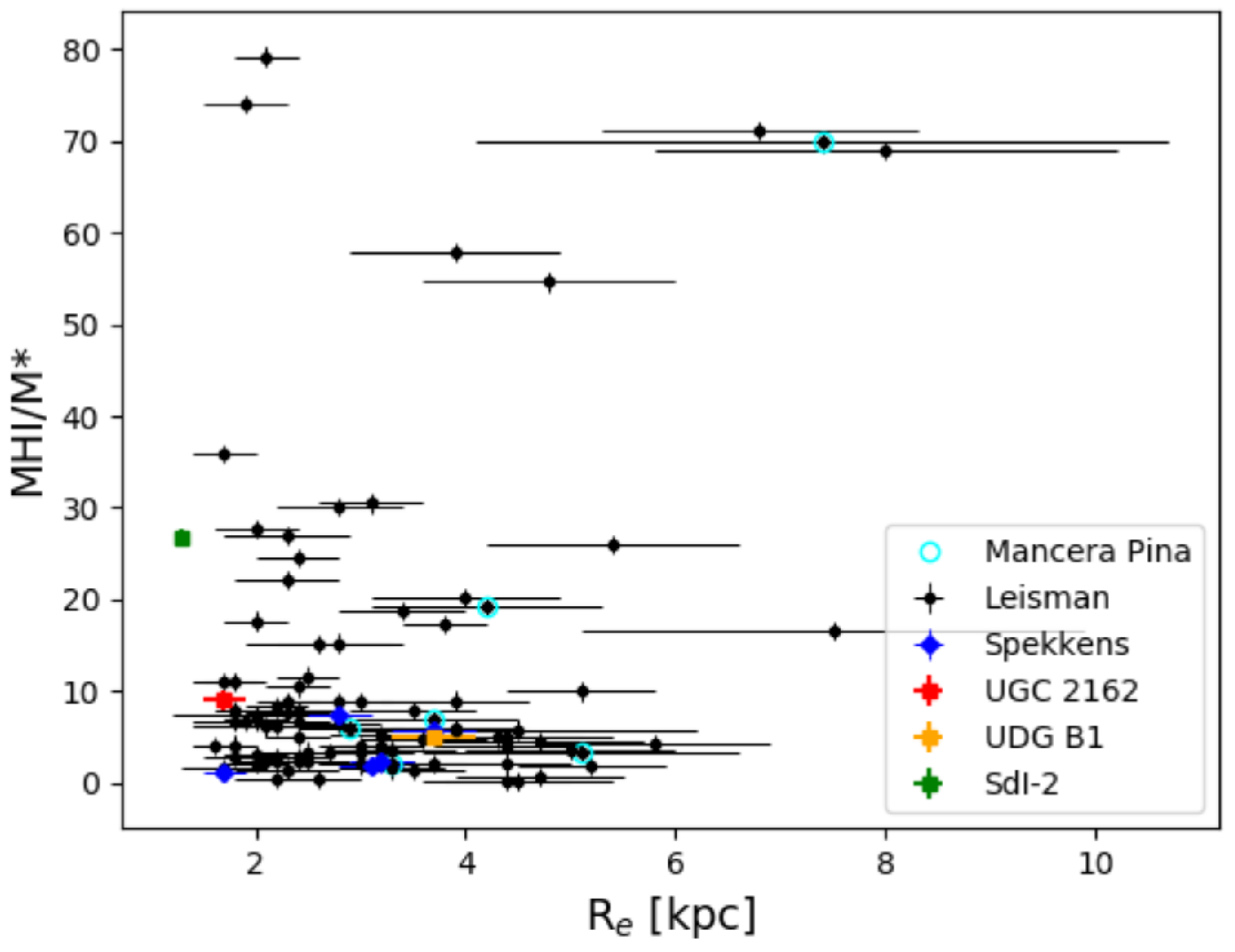}
\vspace{1cm}
\caption{\textbf{$M_{HI}$ / $M_*$ v \re\ for UDG--B1, \sd\ and selected UDGs reported in the literature} i.e. from Mancera Pi\~na et al. (2019, 2020), Leisman, et al., (2017), Spekkens \& Karunakaran (2018), Sengupta et al. (2019.)  }
\label{fig5}
\end{center}
\end{figure*}

Measurent of \sd's   GMRT \hi\ spectrum gives   V$_{\mathrm {HI}}$  = 2565 $\pm$4 \km\ and W$_{20}$ = 80 $\pm$ 6 \km.   V$_{HI}$ = 2543 \km\ and W$_{50}$ = 69 \km\ were reported for the galaxy in \cite{papastergis17}  using observations from the Effelsberg 100 m telescope.
\sd's velocity field in Figure \ref{fig3} (lower right panel) shows two distinct kinematic regions; the main \hi\ body of the galaxy, which has  a length of $\sim$ 74 $^{\prime\prime}$ (14 kpc) and displays \textcolor{black}{a regular velocity gradient indicating a \textcolor{black}{possible} rotating \hi\ disc ($\sim$ 2527 \km\ to 2618 \km) and a morphologically detached  \hi\ extension to the NE with a narrow range of velocities from $\sim$ 2590 \km\ to 2611 \km. \hi\ velocities in the detached region  are similar to  those in NE of the main galaxy body}   The systematic change in PA ($\sim$ 90\degree) of the iso-velocity contours in the main body, indicates the \hi\ disc is significantly warped.  Based on the \hi\ major and minor axes of the main body of the galaxy in medium resolution \hi\ map we estimated the inclination of the \hi\  disc at 67\degree.   Using this inclination and W$_{20}$ from the medium resolution \sd\ cube and  estimated V$_{rot}$ = 41 \km\  using equation \ref{eqn1}.  We also attempted to use   \textsc{bbarolo} to fit a 3D model to the GMRT \hi\ medium resolution cube  which  gave a fit with an  inclination $\sim$ 60\degree and V$_{rot}$ $\sim$ 35 \km, but  the low number of beams across the disc ($\sim$ 3) and low S/N ratio in the  cube meant we could not place a strong reliance on the fit, but the inclination and V$_{rot}$ from the two methods are  in reasonable agreement. The PV diagram (Figure \ref{fig6}) is from  a cut (PA =23.3\degree)  oriented along the \hi\ disc's major axis. The PV diagram is consistent with a bulk of  \hi\ consisting of a rotating  disc with the detached NE region only appearing in a narrow range of velocities ($\sim$ 2590 \km\ to 2611 \km). Figure \ref{fig6} also shows low level emission  extending NE  from  main \hi\ body toward  the detached \hi\  region and indicating the two \hi\ structure are likely to be associated. This is discussed further in section \ref{dis_interaction}.

\nocite{Leisman17}

\begin{figure*}
\begin{center}
\includegraphics[ angle=0,scale=1.0] {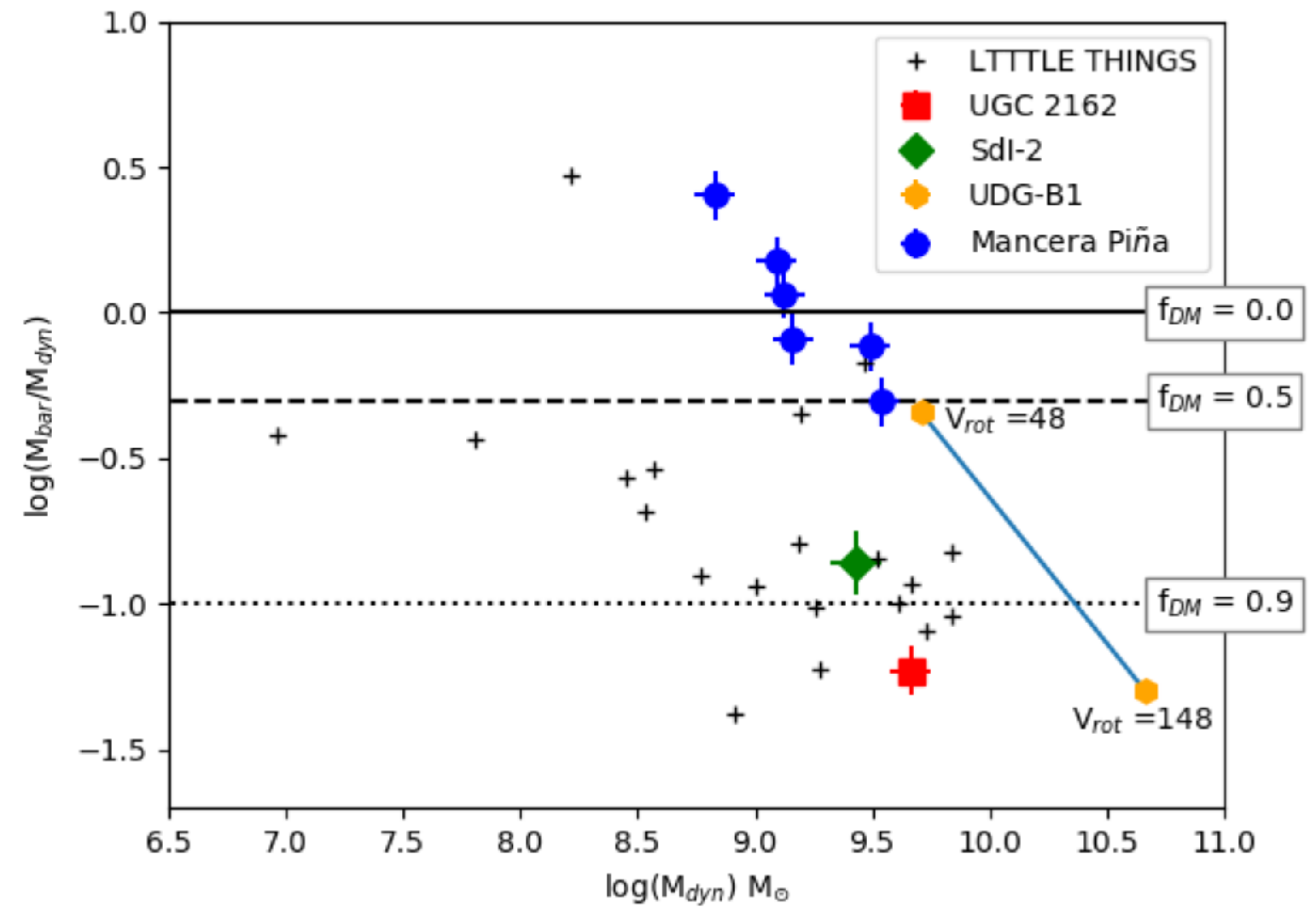}
\vspace{1cm}
\caption{\textbf{$M_{bar}$ / $M_{dyn}$ v $M_{dyn}$:} for UDG--B1 and \sd\ in comparison  to the UDG samples from Mancera Pi\~na  et al., 2020), dwarf galaxies LITTLE THINGS (Oh et al. 2015) and UDG\,2162 (Sengupta et al. 2019.). For UDG\,B1 we show the M$_{dyn}$ determined using the two values V$_{rot}$, 48 \km\ and 148 \km\ discussed in Section \ref{results_kin}. The blue line indicates the range  these two values cover. The horizontal  lines indicate the  dark matter fractions shown in  right hand margin.} 
\label{fig4}
\end{center}
\end{figure*}
\nocite{oh15}

\section{Discussion}
\label{dis}

\subsection{Dark matter  halos }
\label{dis_dyn}

\textcolor{black}{Ideally a UDG's dynamical  mass would be determined using its \hi\ rotation curve which in the field or in groups can probe a galaxy's dynamical mass well beyond the radius of its stellar disk. By fitting a \hi\ rotation curve to DM halo models the mass and mass profile of the DM halo can be estimated. But in clusters because of the absence of \hi\ due to  efficient \hi\ stripping by the cluster environment,  dynamical and DM halo mass estimates to date are typically based on globular cluster studies. \citep[e.g.][]{beasley16, beasley16a, vanDokkum18N,vanDokkum19A,vandokkum19} This approach is supported by the scaling relation between the number of globular clusters and the  host galaxy's halo virial mass \citep{burkert20}, although the impact of the cluster on a UDGs globular population remain unclear \citep{forbes20}.  In the case of the Coma cluster UDG DF--44, which has the largest known DM halo mass of M$_{200}$ =  10$^{11}$ -- 10$^{12}$, stellar kinematics provided the dynamical mass \citep{vandokkum19b}.  There are a small number of Coma UDGs, like DF--44,  which have high numbers of globular clusters per unit stellar mass implying extremely massive DM halos  with two thirds of Coma  UDGs (\re\ $>$ 1.5 kpc) having halo masses $\geq$ 10$^{11}$ \msolar\ based on  globular cluster--DM halo mass scaling relations  \cite {forbes20}.  These halo masses are consistent with the failed Milky Way  mass galaxies. Earlier globular cluster-based studies of galaxy cluster UDGs had indicated that while high dynamical to stellar mass ratios were} common, e.g. VCC\,1287 and DF\,17, typically those studies implied dwarf mass DM halos \citep[e.g.][]{beasley16, beasley16a}. 

An important recent development has been reports  of \textcolor{black}{baryon} \textcolor{black}{dominated} UDGs based on globular cluster studies \citep[][]{vanDokkum18N,vanDokkum19A,vandokkum19} and resolved \hi\ \cite{mancera19,Mancera20}. 
\cite{Zaritsky17}, using scaling relations, suggested that UDGs may span a range of DM halo masses between those typically found in large spirals to dwarf galaxies, but further observations are needed to confirm this.
Deriving an accurate estimate of a UDG's  DM content from \hi\ requires a rotation curve extracted from resolved \hi\  observations. To date  only a few UDGs have been  mapped in \hi\   \citep[e.g.][]{Leisman17,sengupta19,mancera19}.  Interestingly,  several of these UDGs with resolved \hi\  show an apparent departure from the \cite{mcGaugh00} baryonic  Tully Fisher relation (BTFR). Analysis of the  DM content of six of the  UDGs in the Leisman sample of isolated \hi\ detected UDGs   \citep{mancera19,Mancera20} indicates that those galaxies are baryon dominated  within their $R_{HI}$, \textcolor{black}{see Figure 4 in \cite{mancera19},} 
although their results are based on a \textcolor{black}{limited number} of  \textsc{bbarolo} \citep{DiTeodoro15} 3D kinematic tilted--ring disc model fits for each galaxy.  A similar, although lower magnitude,  departure from the BTFR  \textcolor{black}{was} reported in \cite{sengupta19} for the UDG, UGC\,2162 also derived \textcolor{black}{ from \textsc{bbarolo} fitting}.   Confirmation of a baryon dominated  population of UDGs has  important implications for the origin of those UDGs. 
Additionally, the \cite{mancera19,Mancera20} galaxies as well as UGC\,2162  have a lower V$_{rot}$ compared to normal dwarf galaxies of similar \hi\ mass \citep[]{Mancera20, sengupta19}. \textcolor{black}{This is consistent with the results from the larger \cite{Leisman17} sample which found the mean velocity width, W$_{50}$, for UDGs was significantly narrower ($<$44 $>$ \km) than for ALFALFA galaxies with a similar selection criteria $<$119 $>$ \km).}  

\textcolor{black}{As noted in section \ref{results_kin}   the \textsc{bbarolo} 3D kinematic tilted--ring disc models  to the \hi\  cube for UDG--B1 failed and for \sd\ the fit provided only a \textcolor{black}{poorly} constrained estimate the  DM halo properties.} So as an alternative, we estimated their dynamical masses ($M_{dyn}$) from the available  \hi\ properties. We did this using  the \hi\ radii from the medium resolution GMRT integrated \hi\  maps for UDG--B1 and \sd\ ($\sim$ 4.0 kpc and 6.8  kpc respectively) and their inclination corrected  V$_{rot}$ \textcolor{black}{148 \km\ (UDG--B1) and 41 \km\ (\sd)}.  Using this method their respective  $M_{dyn}$ was estimated at \textcolor{black}{45.9} $\times$ 10$^9$ \msolar\ 
and  \textcolor{black}{2.7} $\times$ 10$^9$ \msolar. The baryonic masses ($M_{baryon}$) of the galaxies (\mhi\ $\times $ 1.4 + $M_*$) from Table \ref{table2}  equal 1.76 $\times$ 10$^9$ \msolar\ and 0.35 $\times$ 10$^9$ \msolar\ for  UDG--B1 and \sd, respectively. The 1.4 factor applied to the \hi\ masses was to adjust for the helium and molecular gas content of the UDGs.  Based on these values the baryon fractions for UDG--B1 and \sd\ are  \textcolor{black}{0.05 and 0.14}, respectively. These fractions indicate that, within the radii where \hi\ was detected,  both UDG--B1 and the main body of \sd\ are both dark matter dominated.

However as noted in Section \ref{results_kin}, the UDG--B1  inclination corrected V$_{rot}$ estimate of 148 \km\ derived from from the \hi\ axial ratio  is anomalously high. If  instead we use  V$_{rot}$ = 48 \km\ based on the optical axial ratio it   implies a $M_{dyn}$ = 5.1 $\times$ 10$^9$ \msolar\ and baryon fraction ($M_{bar}$ / $M_{dyn}$) increases to  0.45.   In figure \ref{fig4}, originally from \cite{Mancera20},  we show the log of the ratio of baryonic to dynamical mass versus the log of dynamical mass for UDG--B1, \sd, and UGC\,2162 \citep{sengupta19} as well as the  galaxies from the LITTLE THINGS \citep{oh15}, and \cite{mancera19} as shown in \cite{Mancera20}. For UDG-- B1 we show the masses derived using both V$_{rot}$ = 148 \km\ and 48 \km, with a blue line joining the two sets of derived masses. The horizontal lines indicate the DM fraction (f$_{DM}$)  shown at the right--hand margin.    For UDG--B1 the baryonic (or inversely DM) fraction depends on the adopted V$_{rot}$ which in turn depends on whether  the \hi\ or optical axial ratio is used. Use of the optical axial ratio implies V$_{rot}$ = 48 \km\ and a baryon fraction of 0.45, which is close to that of the  galaxy in the \cite{mancera19} sample with the lowest baryon fraction. Alternatively,  using the \hi\ axes ratio and V$_{rot}$ =148 \km\ gives  much lower baryon fraction of 0.05, which would place it near the extreme of DM dominated dwarfs. As observed in Section  \ref{results_kin}  there are reasons to think the V$_{rot}$ values from which these baryonic fractions are derived are  upper and lower limits. As a result we consider its likely that true baryonic fraction of UDG--B1 lies at an intermediate point in  range 0.05 to 0.45 and thus is likely be similar to the median value for the LITTLE THINGS dwarfs, see Figure \ref{fig4}.    

Hence, in terms of DM \textcolor{black}{content}, our estimates of $M_{dyn}$ for  \sd, and the uncertainly for UDG--B1, suggest the DM halos within the \hi\ radii resemble normal DM dominated dwarf galaxies. Figure \ref{fig5} shows the \mhi/$M_*$ v  \re\ for UDG--B1, \sd\ and  a selection of UDGs from the literature, including the baryon dominated UDGs reported in  \citep{mancera19,Mancera20}, which have \re\ $\gtrsim$ 3 kpc. \re\ and their uncertainties in the plot are from the literarure sources, principally from \cite{Leisman17} which are based on their own photometry of SDSS images.  We note from the Figure that  UDG--B1 has a \re\  within the \re\ range where baryon dominated  UDGs have been reported. \textcolor{black}{This suggests that for UDGs outside clusters their \re\ is determined by factor other than their baryon fraction and $\frac{M_{HI}}{M_*}$ ratio. }

\subsection{Evidence for  ongoing interactions }
\label{dis_interaction}
Figure \ref{fig2} (left middle panel) shows  the  IAC stripe 82 optical image of the UDG--B1 and the two small over plotted yellow circles indicate the  positions and 3 arcsec diameters of the SDSS spectra fibres.   Each spectra is  from a different galaxy  with the spectrum in the south from UDG--B1 and the one above it from a background galaxy.  With the aid of the SDSS spectra and a NUV (GALEX) image (Figure \ref{fig2} -- lower right panel)  we interpret the optical image as showing that UDG--B1 is projected in front of a z=0.69010.$\pm$.0005 background galaxy and  extends in faint blue emission $\sim$ 30 arcsec (12.5 kpc) north of the strong star forming region detected in the  SDSS UDG--B1 spectrum.  This picture  of  UDG--B1's elongated N--S morphology is confirmed by the GALEX NUV emission in Figure \ref{fig2} -- lower left panel.

From the  UDG--B1 GMRT medium and high resolution  \hi\ maps we see  the \textcolor{black}{ high column density \hi\ is located} W of the optical/NUV axes  (Figure \ref{fig2}). UDG--B1's \re\  is 3.7 kpc \citep{roman17} and  the NED major optical axis of 0.35 arcmin indicates the $R_{opt}$ is $\sim$ 4.4 kpc.    Based on the \textcolor{black}{first order} 43.3 arcsec  major axis estimate  from  medium resolution \hi\ map 
 we estimate the $R_{HI}$ at 9 kpc and the $R_{HI}$/ $R_{opt}$ ratio = 2.0.  So even allowing for the asymmetric nature of the \hi\ distribution and uncertain inclination  the UDG--B1 $R_{HI}$/ $R_{opt}$ ratio is consistent with  the typical $R_{HI}$/ $R_{opt}$ of 1.8 for late type galaxies \citep{broeils97}, i.e, we see no clear indication  that the UDG--B1  \hi\ disc is truncated. 

The low resolution \hi\ velocity field for UDG--B1 (Figure \ref{fig2} -- \textcolor{black}{upper} right panel)  shows an overall rotation pattern  running  approximately perpendicular to the optical/NUV  axes, but the change in iso--velocity angle from NE to W suggesting a possible warped disc.  
The \hi/optical axis offset, \hi\ warp and anomalous $V_{rot}$  are all signatures of  an  interaction well within the \hi\ relaxation  time scale of $<$0.7 Gyr \citep{holwerda11}. Such an   interaction is likely to have been with a member of HCG\,25 group.  As noted in Section  \ref{res-morph} the medium and high--resolution \hi\  maps show the current blue star--forming region is offset in projection from the \hi\ column density maxima which could be a further  signature of a recent tidal interaction.  A galaxy's   \af\ ratio is a measure of the asymmetry in its  integrated \hi\ flux density profile  (within its W$_{20}$ velocity range) at velocities above and below the  galaxy's systemic velocity (V$_{\mathrm {HI}}$). \af\ = 1.00 is a perfectly symmetric spectrum and  \af\ $>$ 1.26 presents a clear asymmetry signature \citep{espada11,scott18,Reynolds20}. The \af\ for UDG--B1 is 1.23 $\pm$ 0.11  providing only marginal support for a recent interaction, particularly in because low signal to noise is understood to increase \af\ values \citep{watts20}.

UDG--B1's  SDSS spectrum  cannot be considered representative of the galaxy as a whole because of the 3 arcsec diameter the SDSS spectrum fibre samples only $\sim$ 2.6\% of the UDG--B1 optical \re\ disc area \textcolor{black}{and this region has a bluer color than the rest of the galaxy \citep{roman17}.} Measurements derived from the spectrum, including oxygen abundance, \textcolor{black}{$<$12+log(O/H)$>$ = 8.01$\pm$0.39}, and SFR(\halpha) $\sim$ 0.0033 \msolaryr\  are set out in Table \ref{table3}.  In this table $<$12+log(O/H)$>$ was obtained as the average from the \cite{marino13}, \cite[][lower branch]{kobulnicky99}  and  \cite{petini04} \ calibrators. The starformation rate was calculated assuming the \cite{kennicutt98} conversion formula after correction for Chabrier IMF. \textcolor{black}{The spectrum was corrected for galactic extinction and   internal extinction was insignificant.} The sub--solar 12 +Log(O/H) values  for UDG--B1 indicate that the gas, at least at the position of the spectrum,  is unlikely to have been acquired from more  evolved members of HCG\,25. \textcolor{black}{Additionally, analysis of the SDSS spectra by \cite{roman17} indicates a stellar age $<$ 0.1 Gyr for the region within the SDSS fibre. This young stellar age may simply reflect a local stochastic increase in UDG--B1s star formation, but presence of \hi\ perturbation signatures  make it more likely  the  recent enhancement of SF was triggered by an interaction. }  In summary the  UDG--B1 \hi\ morphology and kinematics both show  indications of a recent interaction, which has perturbed its \hi\ disc, most likely with another member of HCG\,25 and the blue color and young age of the stellar population at the position of the SDSS spectrum is consistent with star formation triggered by the interaction.


\begin{table}%

\begin{minipage}{75mm}

\begin{center}
\caption{\textcolor{black}{UDG-B1: Emission line fluxes and properties from the SDSS spectrum. The average 12 + log(O/H) is obtained from the Marino et al. (2013), Kobulnicky et al. (1999; lower branch) and  Pettini \& Pagel (2004) calibrators, respectively.}}
\label{table3}
\begin{tabular}{lrlr}
\hline
I($\lambda$)/I(H$\beta$) $\times$ 100 \\
\hline
$[$SII$]$6731  & 257.08 $\pm$8.41 \\
$[$SII$]$6717  &  95.64 $\pm$3.85 \\
$[$NII$]$6584  &   3.88 $\pm$0.09 \\
H$\alpha$& 177.78 $\pm$5.59 \\
$[$OIII$]$5007 & 257.08 $\pm$8.41 \\
$[$OIII$]$4959 &  95.64 $\pm$3.85 \\
H$\beta$ & 100.00 $\pm$4.47 \\
$[$OII$]$3727  & 253.08 $\pm$14.54 \\
\hline
F(H$\alpha$) $\times$10 $^{-16}$ erg cm$^{-2}$ s$^{-1}$ & 7.72 $\pm$0.07 \\
F(H$\beta$)  $\times$10 $^{-16}$ erg cm$^{-2}$ s$^{-1}$ & 4.34 $\pm$0.10 \\
H$\alpha$/H$\beta$ $\times$100 & 177.78 $\pm$0.06 \\
c(H$\beta$)        &   0.00 $\pm$0.03 \\
EW(H$\alpha$) $\AA$      & 115.0 \\
EW(H$\beta$)  $\AA$     &  41.0 \\
\hline
log($[$OIII$]$/H$\beta$) &  0.41 $\pm$0.03 \\
log($[$NII$]$/H$\alpha$) & -1.66 $\pm$0.04 \\
log($[$SII$]$/H$\alpha$) & -0.53 $\pm$0.09 \\
\hline
SFR(H$\alpha$) M$_{\sun}$ yr$^{-1}$ & 0.003 $\pm$ 0.001 \\
M(HII) $\times$10$^{5}$  M$_{\sun}$ & 1.71 \\
12+log(O/H)$_{M}$ & 7.98 $\pm$0.19 \\
12+log(O/H)$_{KL}$ & 7.97$\pm$0.26 \\
12+log(O/H)$_{PP}$ & 8.07$\pm$0.22 \\
$<$12+log(O/H)$>$ & 8.01 $\pm$0.39 \\

\hline
\end{tabular}
\end{center}
\end{minipage}{}

\end{table}

  \nocite{marino13}
  \nocite{kobulnicky99}
  \nocite{petini04}

\sd's two most striking features are its high \mhi/$M_*$ ratio of 28.9 \textcolor{black}{(see Figure \ref{fig5})}, which is comparable to the well known extremely gas--rich  dwarf  DDO 154 \citep[\mhi/$M_*$ = 31 ][]{watts18}, and  it's detached  \hi\  \textcolor{black}{ extended NE region  (Figure \ref{fig3}).} \textcolor{black}{Given \sd's isolation, ram pressure stripping can almost certainly be ruled out as an explanation for the \sd\ detached \hi\ extension. So, the most likely explanation for the detached \hi\ extension  is that it  is  debris from the  accretion of a smaller gas--rich satellite}. There are examples in the literature of isolated galaxies with one-sided \hi\ tails and warped discs attributed to the accretion of or interactions with satellite galaxies \citep{matinez09,sengupta12,scott14}.  However, this proposition is only marginally supported by analysis of the integrated GMRT \hi\ spectrum profile, \af\ = 1.20$\pm$0.07.   
While the evidence from the \hi\ profile analysis is inconclusive,  the resolved integrated \hi\  maps and velocity field maps (from the medium resolution cube) provide 
signatures characteristic of a recent interaction. This demonstrates the higher sensitivity to recent \hi\ interactions of resolved mapping compared to \hi\ profile analysis.

Figure \ref{fig3}, lower left panel, shows a smoothed NUV (GALEX) image of \sd. SDSS catalogues the clump marked 'D' in the figure as a star but it could alternatively be a background galaxy. \textcolor{black}{Optically,} \sd\ consists of an elongated region of  bright starforming clumps oriented approxmately N-S.  Three of these clumps, \sd--a, \sd--b\ and \sd--c\  are identified in the Figure  using the nomenclature from \cite{bellazzini17}. These bright clumps are  surrounded by \textcolor{black}{a} larger region of low surface brightness NUV emission \textcolor{black}{(Figure \ref{fig3})}. This low surface brightness emission is more extened in the NE, extending  $\sim$ 20 arcsec (3.6 kpc) NE of the optical centre. \cite{bellazzini17} reported   12+log(O/H) of 8.1$\pm$0.02 for clumps \sd--a and \sd--b, with the difference between them being indistinguishable within the errors. Those authors also reported on a spectrum from the  optical centre, \sd-c, with  model fits indicating a \textcolor{black}{stellar metallicity of Z=0.002 and age of 200 Myr}.  This spectrum differed from \sd--a\ and \sd--b\ in not having any significant emission lines, presumably because of a lack of ionising photons. 
The \sd--c\ log(N/O) from \cite{bellazzini17} is -1.7 implying a primary production of nitrogen in a low metallicity environment. However, the observed uncertainties of N/O at 12 + log(O/H) $\lesssim$ 8.3 has been attributed in the literature to a loss of heavy elements via a strong burst (galactic winds) in the recent past or because of the delayed production of oxygen and nitrogen by massive and intermediate--mass stars assuming a low star formation efficiency. \textcolor{black}{So, a possible explanation of the lack of emission line in the \sd--c spectrum and the NUV halo is a recent strong bust of SF, but the difficultly with this scenario is the absence of the massive star clusters (M* $\sim$ 10$^4$ \msolar) this burst should have produced. We, therefore, conclude that the brightest observed starforming clumps are unlikely to be the result of strong recent star burst.  }
\sd's  \hi\ morphology, together with its   regular  rotating, but wrapped,  main body \hi\ disc   as well as the  perturbed \hi\ kinematics in the NE of the main \hi\ body at similar velocities to those in the detached \hi\ extension (Figure \ref{fig3} Lower right panel) are suggestive of a recent tidal  interaction. While, \sd--a and \sd--b could be remnants of an absorbed  satellite  or star forming regions triggered by such an interaction we do not have definitive proof of this.

\section{Summary and concluding remarks}
\label{concl}
We report on \hi\ mapping using the GMRT for two ultra--diffuse galaxies (UDGs) from contrasting environments. UDG--B1 is projected $\sim$ 225 kpc SW of the compact group HCG\,25 while Secco--dI--2 (\sd) is a relatively isolated UDG. These two UDGs  also have contrasting effective radii with \re\ of  3.7 kpc (similar to the Milky Way) and 1.3 kpc respectively. The \hi\ morphology and kinematics of  UDG--B1 suggest a recent interaction has \textcolor{black}{perturbed both the morphology and kinematics of its  \hi}. UDG--B1's subsolar metallicity suggests it did not acquire a significant gas mass in the interaction.    \sd\ has two striking features, first its  $\frac{M_{HI}}{M_*}$ ratio  is 28.9  and despite its isolation, it displays a  one--sided detached extended \hi\ region to the NW. It's $\frac{M_{HI}}{M_*}$ ratio implies a very low historic star formation efficiency. Based on our estimate of $M_{dyn}$ the baryon fraction within  \hi\  radius for \sd\  is  0.14,   indicating that  galaxy is dark matter dominated, at least within the radius in which \hi\ is detected. However, the baryon fraction for UDG-B1 is quite uncertain because of large uncertainties about the \hi\  disc's inclination and the question as to whether its \hi\ is in dynamical equilibrium. As a result we were unable to reliably estimate the UDG--B1 DM fraction.  The higher  \mhi/$M_*$ ratio in \sd\ compared to UDG--B1 is consistent with the result from \cite{papastergis17}  which indicated UDGs further from group centres have higher \mhi/$M_*$ ratios. \textcolor{black}{In the case of UDG--B1 morphological evidence of  tidal disturbance and possible stripping (and possibly ram pressure stripping)  could explain the smaller stellar mass and redder colour of UDGs at distances closer to group centres as reported in \cite{papastergis17}}. Our study  highlights the importance of   high spatial and spectral resolution \hi\ observations for the study of the dark matter properties of UDGs. While the narrow  range of \hi\ velocities in UDGs argues for a velocity resolution below $\sim$ 10 \km,  going beyond this level also requires good signal to noise, which adds a further burden to already demanding observations.

\label{summary}

\section*{Acknowledgements} 
We thank the staff of the {\it GMRT} who have made these observations possible. The {\it GMRT} is operated by the National Centre for Radio Astrophysics of the Tata Institute of Fundamental Research.  TS  acknowledge support by Funda\c{c}\~{a}o para a Ci\^{e}ncia e a Tecnologia (FCT) through national funds (UID/FIS/04434/2013), FCT/MCTES through national funds (PIDDAC) by this grant UID/FIS/04434/2019 and by FEDER through COMPETE2020 (POCI-01-0145-FEDER-007672). TS also acknowledges the support by the fellowship SFRH/BPD/103385/2014 funded by the FCT (Portugal) and POPH/FSE (EC). TS additionally acknowledges support from DL 57/2016/CP1364/CT0009 from The Centro de Astrof\'{i}sica da Universidade do Porto.  \textcolor{black}{PL is supported by a work contract DL 57/2016/CP1364/CT0010 funded by the FCT}. This work was supported by FCT/MCTES through national funds (PIDDAC) by this grant PTDC/FIS-AST/29245/2017. Support for this work was provided by the National Research Foundation of Korea to the Center for Galaxy Evolution Research (No. 2010-0027910) and NRF grant No. 2018R1D1A1B07048314. This research has made use of the NASA/IPAC Extragalactic Database (NED) which is operated by the Jet Propulsion Laboratory,  California Institute of Technology, under contract with the National Aeronautics and Space Administration. This research has made use of the Sloan Digital Sky Survey (SDSS). Funding for the SDSS and SDSS-II has been provided by the Alfred P. Sloan Foundation, the Participating Institutions, the National Science Foundation, the U.S. Department of Energy, the National Aeronautics and Space Administration, the Japanese Monbukagakusho, the Max Planck Society, and the Higher Education Funding Council for England. The SDSS Web Site is http://www.sdss.org/. This research made use of APLpy, an open-source Python plotting package hosted at http://aplpy.github.com,  \citep{Robitaille2012}.

\section*{Data Availability } 
The data underlying this article will be shared on reasonable request to the corresponding author.
\bibliographystyle{mnras}
\bibliography{cig}

\begin{figure}
\begin{center}
\includegraphics[ angle=0,scale=.4] {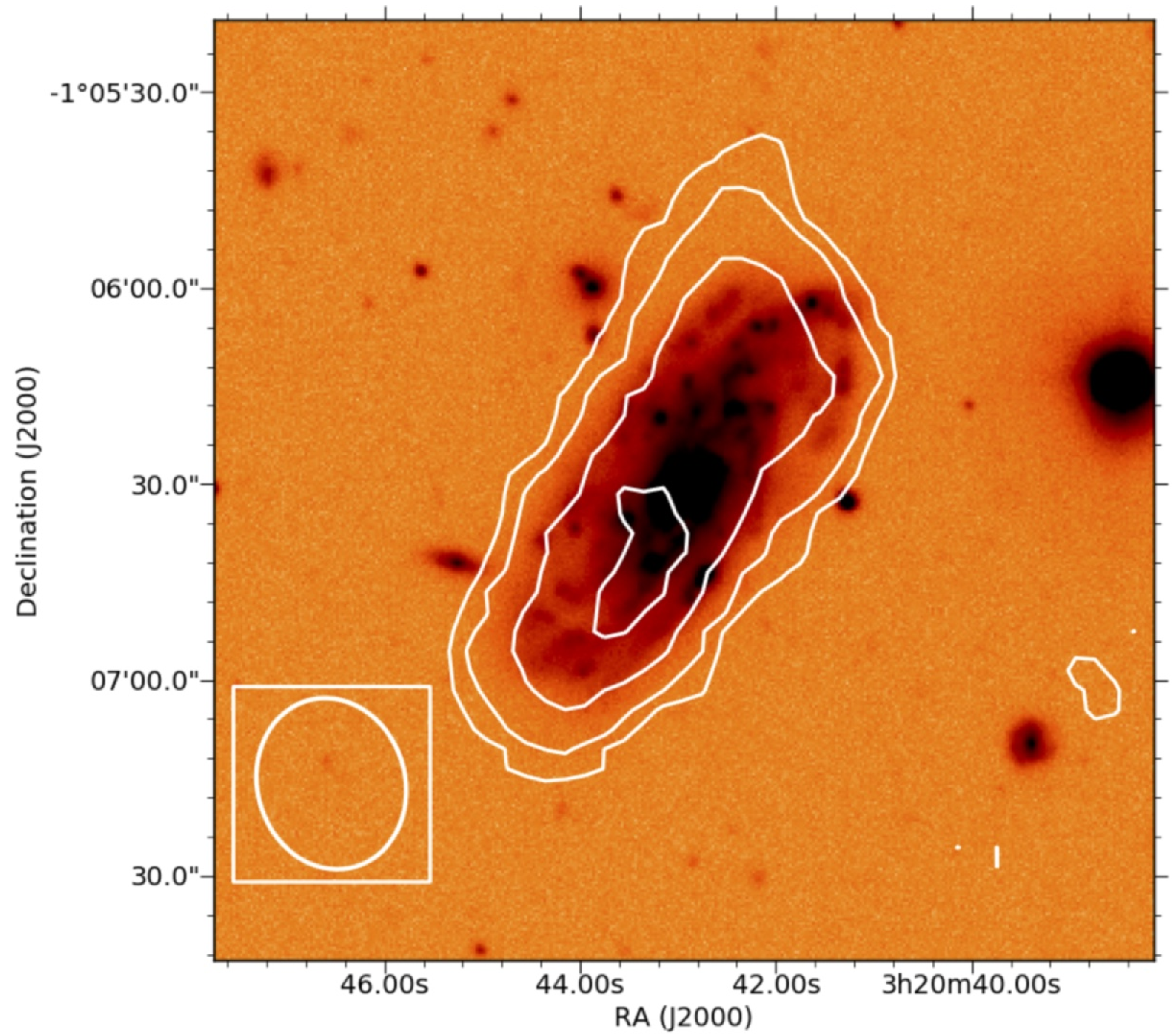}
\vspace{1cm}
\caption{\textbf{ UGC\,2690: }GMRT medium resolution  (26.46$^{\prime\prime}$ $\times$ 22.65$^{\prime\prime}$) \hi\ velocity integrated map contours on Stripe 82 SDSS g--band image. The contours are at  2.3, 4.6, 9.2 and 13.8   $\times$ 10$^{20}$ atoms cm$^{-2}$. The GMRT beam size and orientation  is shown with white ellipse at the bottom left of the figure.}
\label{fig2690}
\end{center}
\end{figure}

\begin{figure*}
\begin{center}
\includegraphics[ angle=0,scale=.50] {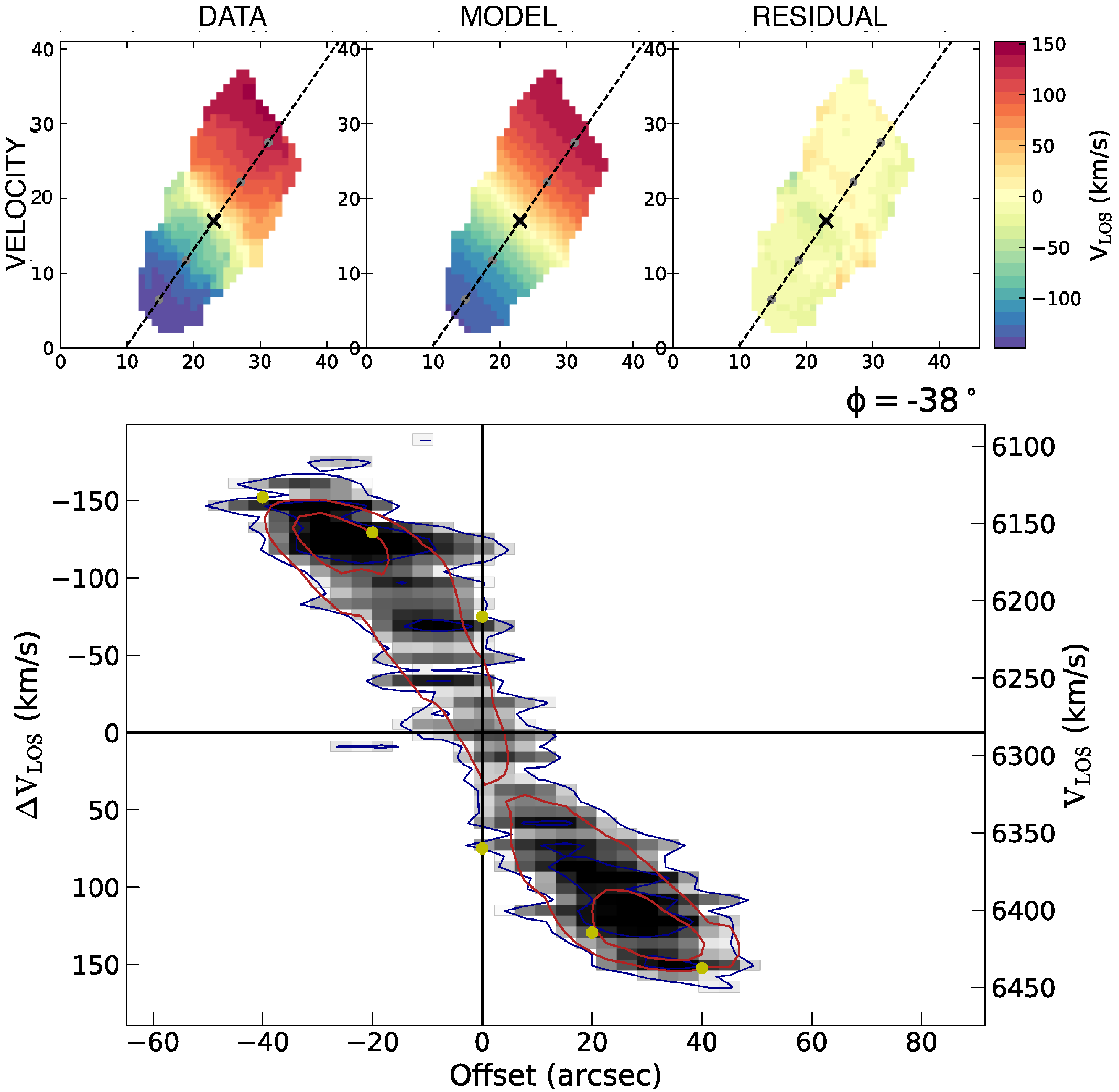}
\vspace{1cm}
\caption{\textbf{ UGC\,2690: } \textbf{top:} \textsc{bbarolo} data, model and residual \textcolor{black}{from}  the medium resolution  (26.46$^{\prime\prime}$ $\times$ 22.65$^{\prime\prime}$) GMRT \hi\ cube. \textbf{Bottom:} \textsc{bbarolo} PV diagram for PA =128\degree\ from the same cube. The blue contours are from the data and the red contours are from the model, with the yellow dots showing the fitted rings. }
\label{fig2690BB}
\end{center}
\end{figure*}

\newpage
\noindent\textbf{APPENDIX A}\\
\bigskip
\textbf{UGC\,2690}\\
\hi\ was also detected in  the SAc galaxy UGC\,2690, within GMRT UDG B1 \textcolor{black}{field of view},   at V$_{\mathrm {HI}}$ = 6275$\pm$ 4 \km\ with W$_{20}$ 340$\pm$ 8 \km. The \hi\ morphology  from the medium resolution cube  is shown in Figure \ref{fig2690} and displays an asymmetric increase in minor axis diameter NE of the optical centre. A similar asymmetry is seen in optical image.  It seems likely both the \hi\ and optical  features are  due to a recent interaction with another group member. The V$_{\mathrm {HI}}$  = 6269$\pm$ 2 \km\ and W$_{20}$ = 340$\pm$ 8 \km\ from the GBT spectrum \citep{springob05}  agree within the uncertainties to the GMRT values.  Measured from the GMRT spectrum the  \af\ = 1.22$\pm$0.09 and \dvs, as defined by \cite{Reynolds20}, = 5.3$\pm$6. As \textcolor{black}{was the case} for SdI--2 the resolved mapping is more sensitive to indications of perturbation  than the \hi\ profile parameters. Figure \ref{fig2690BB} shows the \textsc{bbarolo} velocity field, data, model and residuals from the medium resolution cube and the PV diagram (PA =128\degree) for UGC\,2690. The  \textsc{bbarolo} model fit indicates  an \hi\  V$_{rot} $ = 150 \km\ $\sim$ inclination of $\sim$ 79\degree.

\end{document}